\documentclass[aps,eqsecnum]{revtex4}
\usepackage{graphicx,epsfig}
\begin{document}
\def\ii{\'{\i}}
\title{Pentaquarks in the medium in the quark-meson coupling model}

\author{P.K. Panda}
\affiliation{Centro de F\ii sica Te\'orica - Dep. de F\ii sica -
Universidade de Coimbra - P-3004 - 516 Coimbra - Portugal}
\author{C. Provid\^encia}
\affiliation{Centro de F\ii sica Te\'orica - Dep. de F\ii sica -
Universidade de Coimbra - P-3004 - 516 Coimbra - Portugal}
\author{D.P. Menezes}
\affiliation{Depto de F\ii sica - CFM - Universidade Federal de Santa
Catarina  Florian\'opolis - SC - CP. 476 - CEP 88.040 - 900 - Brazil}
\affiliation{School of Physics - University of Sydney - NSW - 2006 - Australia}

\begin{abstract}
We calculate the properties of the pentaquarks $\Theta^+$ and
$\Xi^{--,0}$ in  symmetric nuclear matter using the quark meson
coupling model (QMC). The stability of the $\Theta^+$ in the medium
with respect to the channel $\Theta^+\to NK^+$  is discussed.
\end{abstract}
\date{\today}
\maketitle
\vspace{0.5cm}
PACS number(s): 36.10.-k,24.10.Jv, 21.65.+f
\vspace{0.5cm}

\section{Introduction}

The possible existence of pentaquarks was first proposed a long time ago
\cite{old} but the subject regained theoretical interest after such exotic
states were indeed measured \cite{leps,outros}. In the recent months many
works have been dedicated to the calculation of pentaquark properties
\cite{stancu,shen,zhong,koreans,marina}, a theoretical pioneer work being
the one that classified a narrow strange pentaquark with positive parity
as the lowest state of an $SU(3)$ decuplet within a chiral soliton model
\cite{diakonov}.

Three of the possible pentaquarks are the $\Theta^+$
\cite{leps,outros} containing the quarks $uudd \bar s$ with mass
1540 MeV, $S=+1$, zero isospin ($I=0$) and positive charge, the
$\Xi^{--}$, formed by $dss d\bar u$, $S=-2$, $I=3/2$
 and the $\Xi^0$ constituted of quarks $dss u\bar d$,
$S=-2$, $I=3/2$ and neutral, the last two with masses equal to 1862
MeV, detected in \cite{casc} but not confirmed in other experiment
\cite{cascoutros}.

Another topic of great interest in the recent literature is the
in-medium modification of baryon properties, mainly their
masses and decay widths \cite{hatsuda85,taps,others,AM}.  

In the present work we use the quark-meson coupling model (QMC)
\cite{guichon,st}
in order to obtain the effective masses of the pentaquarks $\Theta^+$ and
$\Xi^{--,0}$ in a self-consistent way with the nucleon mass. We  fix the
nucleon radius and determine the pentaquark radius considering
the same bag constant of the nucleon.
We compare the chemical potential of the pentaquarks determined
in this way with the chemical potential of the nucleon plus kaon system 
as a function of the density. A comparison with the results obtained
within the  non-linear Walecka model (NLWM) \cite{walecka} is drawn.

\section{The quark-meson coupling model for pentaquarks}

In the QMC model, the baryons are assumed to be static spherical
MIT bags in which quarks interact with the scalar and vector fields,
$\sigma$, $\omega$ and $\rho$ and these fields are treated as
classical fields in the mean field approximation. In the present work we treat
nucleons and  pentaquarks within the same formalism, the differences arising 
from the number and flavor of the constituent quarks. We consider
both the $\Theta^+$ and the $\Xi^{--,0}$ in nuclear matter and assume that
non-strange mesons do not couple with the $s, \, \bar s$ quarks.

The quark field,
$\psi_q(x)$, inside the bag satisfies the equation of motion:
\begin{equation}
\left[i\,\rlap{/}\partial-(m_q^0-g_\sigma^q\, \sigma_0)
-\gamma^0~(g_\omega^q\, \omega_0\, + \frac{1}{2} g^q_\rho \tau_{3q}
b_{03})\right] \,\psi_q(x)=0\ , \quad  q=u,d,s, \label{eq-motion}
\end{equation}
where $\sigma_0$, $\omega_0$ and $b_{03}$ are the classical meson
fields for $\sigma$, $\omega$ and $\rho$ mesons. $m_q^0$ is the current quark
mass, $\tau_{3q}$ is the third component of the Pauli matrices and
$g_\sigma^q$, $g_\omega^q$ and $g_\rho^q$ are the quark couplings
with $\sigma$, $\omega$ and $\rho$ mesons.

The quark energy is
\begin{equation}
\epsilon_q=\Omega_q +R_B\left(g_\omega^q\, \omega+ \frac{1}{2}
g^q_\rho \tau_{3q} \rho_{03} \right)
,
\end{equation}
with $\Omega_q\equiv \sqrt{x_q^2+(R_B\, m_q^*)^2}$,
$m_q^*=m_q^0-g_\sigma^q\, \sigma$, $R_B$ is the bag radius of the
baryon. The energy of a static bag describing the baryon $B$, can be
expressed as
\begin{equation}
E^{\rm bag}_B=\sum_q n_q \, {\Omega_q\over R_B}-{Z_B\over R_B}
+{4\over 3}\,  \pi \, R_B^3\,  B_B\ ,
\label{ebag}
\end{equation}
where $Z_B$ is a parameter which accounts for zero-point motion and
$B_B$ is the bag constant. The set of parameters used in the present
work is given in \cite{pmp} for the bag value $B_B^{1/4}=210.854$
MeV, $m_u^0=m_d^0=5.5$ MeV and $m_s^0=150$ MeV. The effective mass
of the nucleon and the  pentaquark at rest is taken to be
\begin{equation}
M_B^*=E_B^{\rm bag}.
\label{eff-mn}
\end{equation}
The equilibrium condition for the bag is obtained by minimizing the
effective masses $M_B^*$ with respect to the bag radius
\begin{equation}
\frac{\partial M_B^*}{\partial R_B^*} = 0.
 \label{balance}
\end{equation}
All the above expressions are valid for the nucleons and the pentaquarks. We 
have considered the same bag constant $B_B$ for both. The radius and the 
$Z_B$ parameter for the pentaquark are then
obtained from equations (\ref{eff-mn}) (for the vacuum) and (\ref{balance}).

It is not clear if the spin parity of the pentaquark $\Theta^+$ is
$J^P=1/2^+$ or  $J^P=1/2^-$. Constituent quark models predict the
first value \cite{stancu} and chiral symmetric models the second \cite{diakonov}. 
Within the QMC
model, we may have a positive parity state if one of the quarks
occupies a $p$ state. However, this is not the lowest energy state.
In the following we also consider this possibility.

 For simplicity, we assume that the bag describing the nucleon
remains in a spherical shape with radius $R_B$. The
single-particle  energies in units of $R^{-1}$ are given as
\cite{panda}
\begin{equation}
\epsilon_q=\Omega_q^{n\kappa} \pm R_B\left(g_\omega^q\, \omega+ \frac{1}{2}
g^q_\rho \tau_{3q} \rho_{03} \right),
\end{equation}
for quarks and  antiquarks, with
$
\Omega^{n\kappa}_q = \sqrt{x^2_{n\kappa}+R_B^2{m^*_q}^2} .$
The eigenvalues $x_{n\kappa}$ for the state characterized by $n$ and $\kappa$
are determined by the boundary condition at the bag surface,
$
i\gamma\cdot n \psi_q^{n\kappa}=\psi_q^{n\kappa} .
$
We get for the total energy
\begin{equation}
E^{\rm bag}_B=\sum_{q}\frac{\Omega^{n\kappa}_q}{R_B}-{Z_B\over R_B}
+{4\over 3}\,  \pi \, R_B^3\,  B_B\ .
\label{ebag1}
\end{equation}
For the QMC model the equations of motion for the meson
fields in uniform static matter are the ones usually given in the literature
\cite{guichon,st,pmp}. For the necessary self-consistent treatment, one
has to determine
\begin{equation}
\frac{\partial M_B^*(\sigma)}{\partial \sigma} = \frac{\partial
E^{\rm bag}_B}{\partial \sigma} .
\end{equation}

Notice that, although we have included the $\rho$ meson in the equations
above, we shall restrict ourselves to the calculation of symmetric nuclear
matter. Moreover, as we have opted to use identical masses for the $u$ and
$d$ quarks, the pentaquarks $\Xi^0$ and $\Xi^{--}$ are
indistinguishable.

Boosting the bags we get the dispersion relation for the
pentaquarks
\begin{equation}
\mu_\Theta=\sqrt{{M_B^*}^2+\mathbf{k}^2} +g_{\omega}\omega_0.
\label{mutheta}
\end{equation}
We also calculate the optical potential of a pentaquark in symmetric
nuclear matter, which we define as
$$ V_B=M_B^*-M_B+ g_{\omega B}\omega_0.$$

In the numerical calculations we have used \cite{pmp}
$g_\sigma^q=5.957$, $g_{\sigma N}=3g_\sigma^q S_N(0)=8.58$,
$g_{\omega N}=8.981$, $g_{\rho N}=8.651$ with
$g_{\omega N} = 3g_\omega^q$ and $g_{\rho N} = g_\rho^q$, with the standard
values for the meson masses,
$m_\sigma=550$ MeV, $m_\omega=783$ MeV and $m_\rho=770$ MeV.  Note
that in our approach the $s$-quark is unaffected by the $\sigma$, $\omega$
and $\rho$ mesons i.e. $g_\sigma^s=g_\omega^s=g_\rho^s=0\ .$
The pentaquark couplings are given by
$g_{\omega \Theta}=\frac{4}{3}g_{\omega N}$, $g_{\rho \Theta}=g_{\rho N}$.
In Table \ref{tab1} we give the values of $Z$ for both $\Theta^+$  and
$\Xi^{--,0}$ considering the radius of the nucleon 0.6 and 1.0 fm and the
three different cases: a) all quarks in a $s$ state; b) quarks $u$ or
$d$ in a $p$ state; c) quarks $\bar s$ or $s$ in a $p$ state.

In the next section we compare the QMC results obtained in the present work 
with results obtained from the NLWM model \cite{walecka} with the NL3 
parametrization \cite{nl3}. In this model the effective mass of a baryon is
$M^*_B=M_B-g^B_\sigma \sigma$ \cite{glen00}.
Using quark counting arguments and considering that the
$\sigma,\omega$ mesons do not couple to the $s$-quark we take
$$g_{i\Theta}=\frac{4}{3}g_{iN},  \qquad
g_{i\Xi}=g_{iN},\quad i=\sigma,\,\omega.   $$

\section{Results}

The in-medium results obtained with the QMC
model for the pentaquarks are next displayed and compared with the ones 
obtained with the NLWM.

\begin{figure}
\epsfig{file=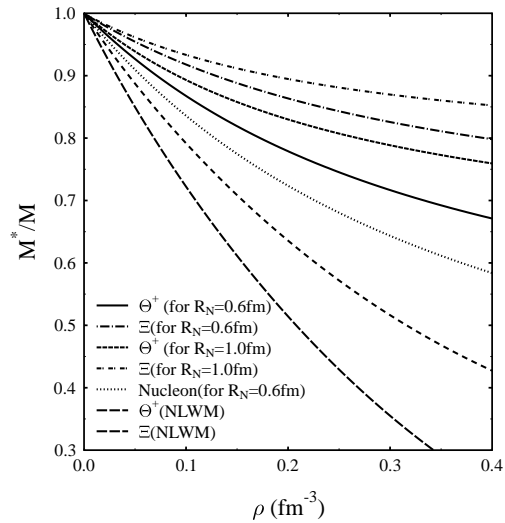,width=7.cm}
\caption{Effective masses of the pentaquarks $\Theta^+$ and $\Xi^{--,0}$ 
within the
  QMC and the NLWM models.}
\label{effm}
\end{figure}

In figure \ref{effm} we show the effective masses for the $\Theta^+$ and the
$\Xi^{--,0}$ for two different values of the bag radius. For the larger value of
the bag radius, the decrease of the effective masses is less pronounced for
both pentaquarks. Our results do not agree with the ones
calculated in \cite{koreans}, where , however,  the nucleon
saturation data of \cite{st} are also not reproduced.

The $\Theta^+$ effective masses always decrease more with density than 
$\Xi^{--,0}$, either within the QMC or the NLWM framework. This is due to the
fact that the $s$-quark does not couple to the $\sigma$-meson and 
$\Xi^{--,0}$ have two $s$-quarks.
For each pentaquark, one can see that the effective
masses always decrease more within the NLWM than within the QMC
model. This effect was already known for the effective mass of the
nucleon: for the the saturation density, QMC gives an
effective mass $\sim$ 0.8 $M_N$ while the NLWM predicts  $\sim$ 0.7 $M_N$.

\begin{figure}
\begin{tabular}{ccc}
\epsfig{file=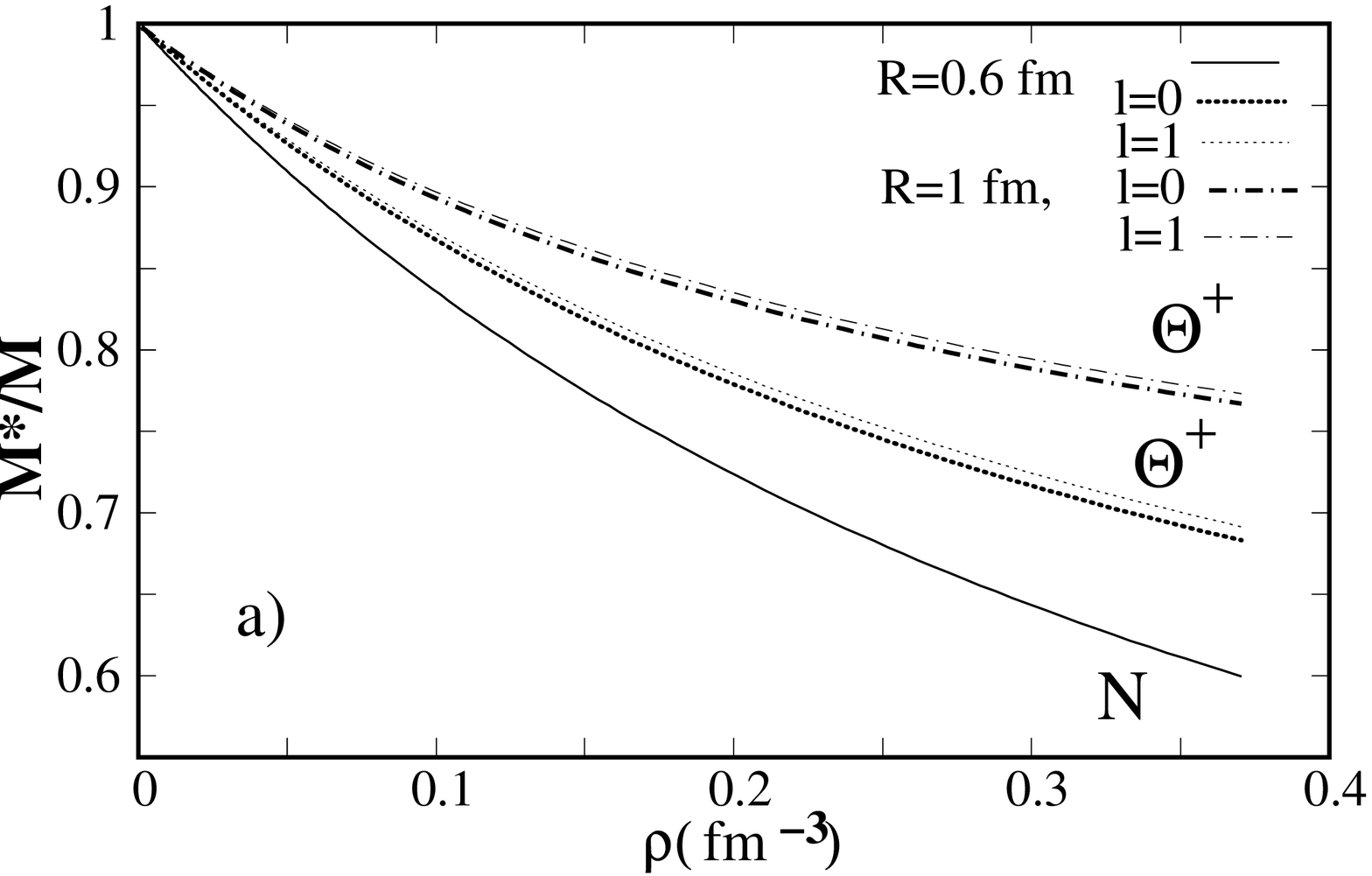,width=7.cm} & 
\epsfig{file=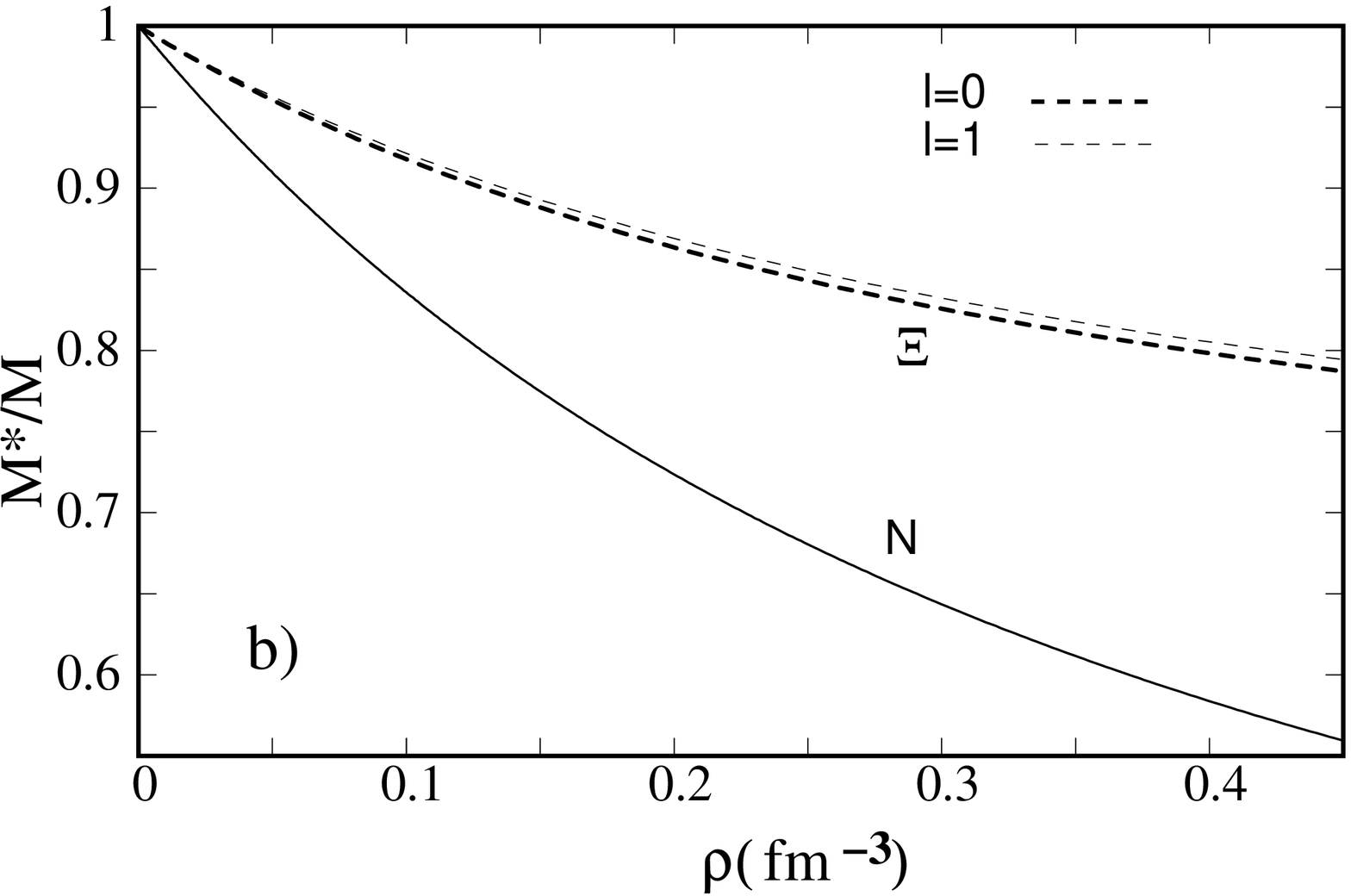,width=7.cm}\\
\end{tabular}
\caption{Effective masses of the pentaquarks $\Theta^+$ and $\Xi^{--,0}$}
\label{meff1}
\end{figure}

In Fig. \ref{meff1} we replot the $\Theta^+$ and $\Xi^{--,0}$ effective
masses in separate graphs. We include the results obtained considering
a quark in a $p_{3/2}$ state. A smaller decrease of the effective mass
with density is observed, but  the mass difference with respect to the
solution with all quarks in a $s$ state  is small.

We have also verified that isospin asymmetry of nuclear matter has little effect on the effective mass of the pentaquarks: for a proton fraction $y_p=0.1$ there is a small increase of the effective mass with density with respect to the symmetric matter value. However,  at 3$\rho_0$ the difference  is only $\sim 1\%$ the symmetric matter value.

\begin{figure}
\epsfig{file=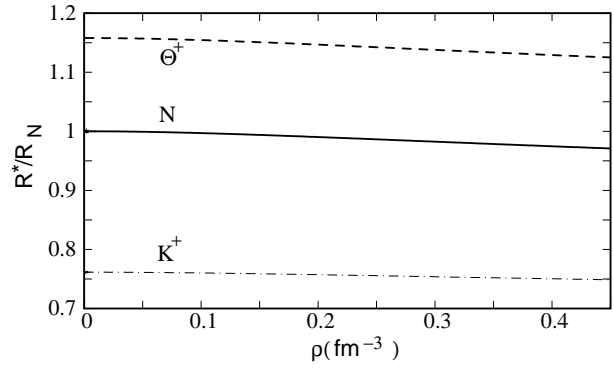,width=8.cm}
\caption{Radius of the $\Theta^+$, nucleon and kaon as a function of density.}
\label{rad}
\end{figure}

We have calculated the radii for three different particles, namely, the 
nucleon, the kaon and the pentaquark $\Theta^+$ fixing the nucleon
radius as 0.6 fm and choosing for the other particles the same bag constant.
It is known that QMC predicts a small reduction of the baryon radius with 
density \cite{guichon96}. In Fig. \ref{rad} we plot, in units of the vacuum
nucleon radius, the radius of these three particles. One can see that the 
$\Theta^+$ radius is only 16\% larger than the nucleon radius and presents the 
same behavior with density.

\begin{figure}
\begin{tabular}{cc}
\epsfig{file=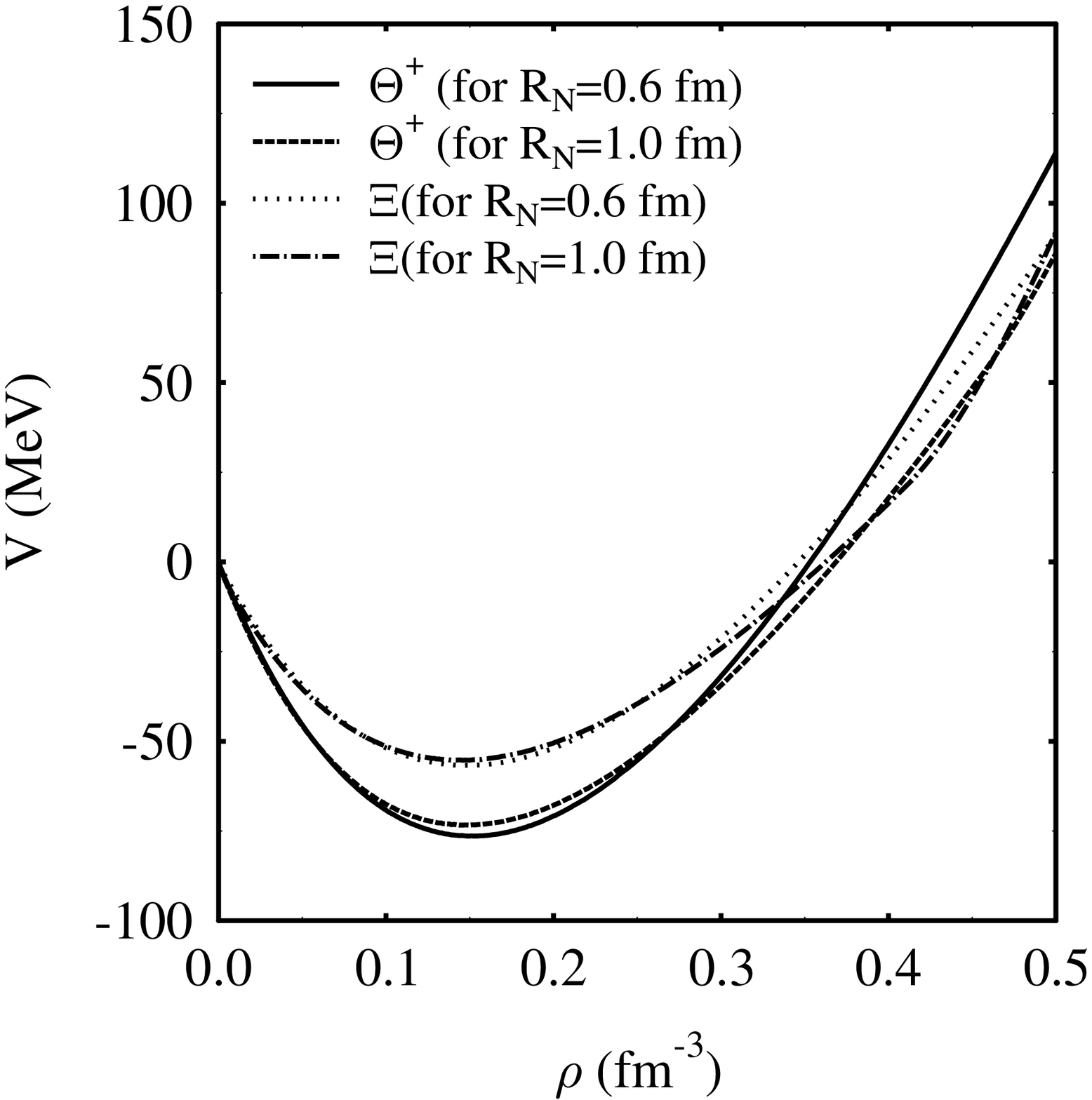,width=7.cm} &
\epsfig{file=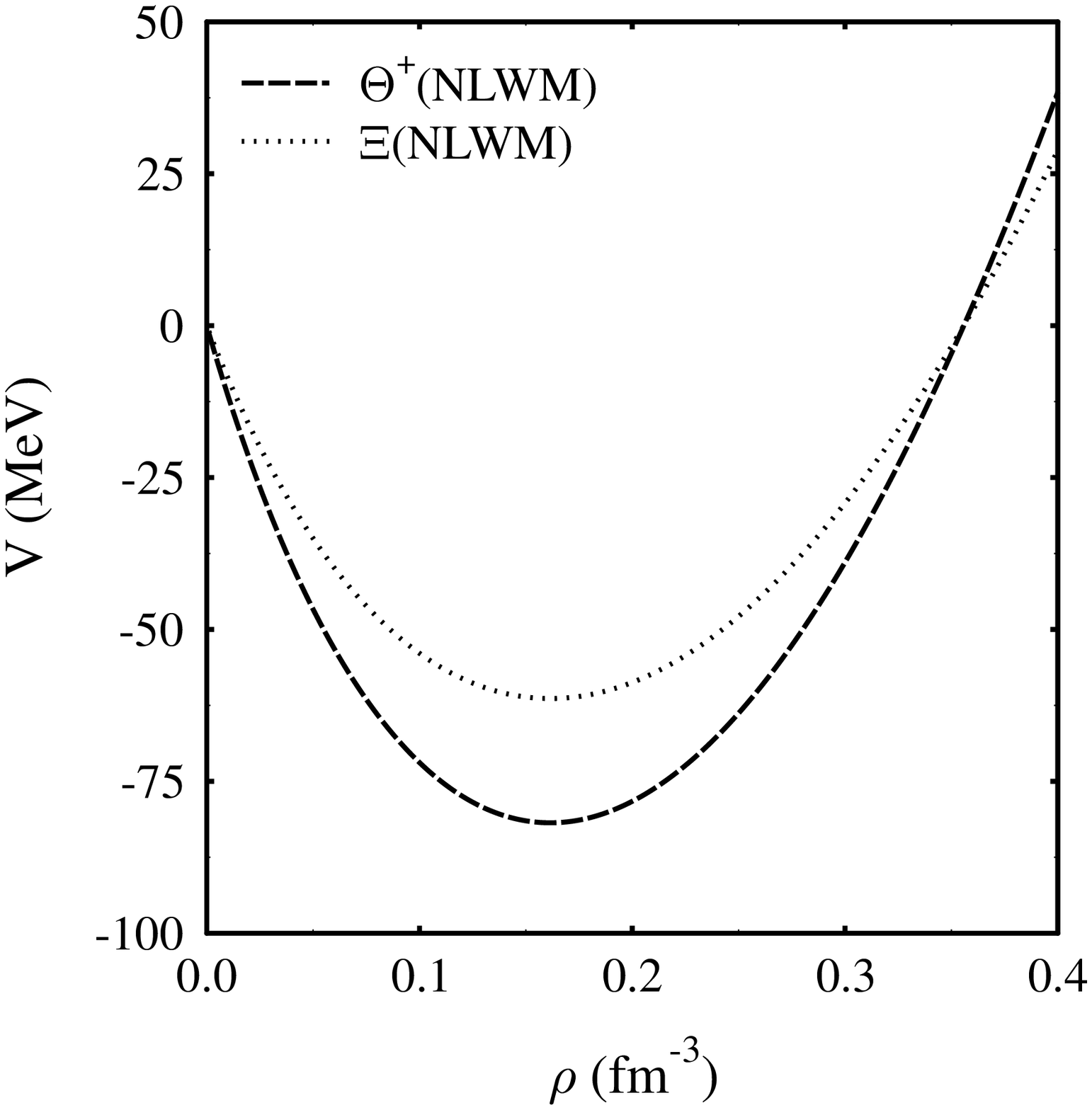,width=7.cm}\\
\end{tabular}
\caption{Optical potentials obtained with the QMC and the NLWM.}
\label{disp}
\end{figure}

In figure \ref{disp} we have shown the optical potentials obtained within
both models discussed in the present work. Within the QMC model we get an
optical potential of $\sim$ $-75$ MeV for the  $\Theta^+$ pentaquark
and  $\sim$ $-50$ MeV for the $\Xi$ pentaquarks. These values are
just a bit smaller than the ones predicted with the NLWM  and are
compatible with values obtained in \cite{oset}, $\sim$ $-70$ to
$-120$ MeV obtained from the calculus of the $\Theta^+$ self-energy tied to a 
two-meson cloud. In \cite{marina} a value $-90 \mbox{MeV}\le U\le -40$ MeV was 
obtained within a QCD sum rule calculation and similar numbers were obtained 
within a relativistic  mean field calculation in \cite{zhong}. Taking
different values for the bag radius has a negligible effect on the potential.

\begin{figure}
\epsfig{file=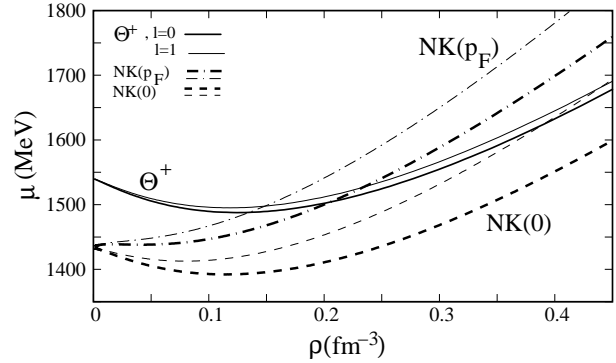,width=8.cm}
\caption{Chemical potentials of the $\Theta^+$, and the $NK^+$ system.}
\label{chem}
\end{figure}

In order to understand under which conditions  the $\Theta^+$ is stable
in the medium with respect to the channel  $NK$ we have calculated
within the QMC model the chemical potential of the pentaquark given by
Eq. (\ref{mutheta}), with the chemical potential of the system nucleon
plus kaon. For the kaon we use the results of \cite{mpp}. We consider
a pentaquark in its rest frame and suppose it decays, $\Theta^+\to
N+K^+$. In the matter, due to the Pauli principle the nucleon must
have a momentum equal to the Fermi momentum. The kaon will have the
same momentum if we don't take into account the effect of the medium.
Taking into account the medium, momentum may
be shared by any other nucleons and the kaon will come out with any
momentum. In Fig. \ref{chem} we plot the chemical potential of the
$\Theta^+$ at rest, and the chemical potential of the system $NK^+$ in
two different cases: a)
both particles with $p=p_F$; b) the nucleon with $p=p_F$
and the kaon with a zero momentum.
Considering the second possibility the pentaquark $\Theta^+$ will
 not be stable in matter. If we consider the first possibility the
 pentaquark $\Theta^+$ is stable for densities $>1.5 \rho_0$.

However, according
to \cite{gqli}, the $K^+$ feels a weak repulsive potential and the
$K^-$ a strong attractive one.  In
order to obtain a value approximately equal to 10 MeV for the $K^+$
potential in the present formalism, 
we must rescaled the coupling $g_{\omega K}$ as 1.625 $g_{\omega
K}$ for the nucleon bag radius $R_N=0.6$ fm and the kaon bag radius
$R_K=0.457$ fm (see \cite{mpp,tsushima98}. Taking this value we get
different results for the chemical potential of the system nucleon plus
kaon represented by thin dashed ($p_K=0$) and dashed-dot ($p_K=p_F$)
lines in Fig.  \ref{chem}. In this case we could have stable
pentaquarks, respectively, for densities larger than $\rho_0$ ($2.5
\rho_0$) if we consider the process $\Theta^+\to
N+K^+$ ($\Theta^+ N\to
N+N+K^+$).

In this short note we have calculated the behavior of the $\Theta^+$
and $\Xi^{--,0}$ pentaquarks in the medium according to the
predictions of the QMC model. In particular we have calculated the
effective mass and the optical potential of those particles as a
function of density. It was shown, by comparing the chemical
potentials of the $\Theta^+$ and the $NK^+$ system that $\Theta^+$
is stable against the decay $\Theta^+\to NK^+$ for  $\rho>1.5
\rho_0$. However, taking into account the medium,  $\Theta^+$ is not
stable even at high densities.
However, these limits are sensitive to the kaon optical potential and
  they   could be lowered, respectively, for $\rho_0$ and $2.5 \rho_0$ if we choose a $g_{K\omega}$ coupling constant
  which reproduces the $K^+$ optical potencial accepted value of 10
  MeV at saturation density. 

 There have been recent calculations of
the pentaquarks in the medium using the same model
\cite{zhong,koreans}. Notice, however, that in \cite{zhong} the properties of 
the $\Theta^+$ have not been calculated in a self consistent way in the QMC
model and in \cite{koreans} the data obtained for the nucleon
properties are not in accordance with the data in the literature 
\cite{guichon,st,guichon96}.

\section*{ACKNOWLEDGMENTS}

This work was partially supported by Capes (Brazil) under process BEX
1681/04-4, CAPES (Brazil)/GRICES (Portugal) under project 100/03 and
FEDER/FCT (Portugal) under  projects  
POCTI/FP/FNU/50326/2003 and POCTI/FIS/451/94.
D.P.M. would like to thank the friendly atmosphere at the Reserch Centre
for Theoretical Astrophysics, Sydney University, where this
work was partially done.

\begin{table}[ht]
\caption{ Bag constant and $Z_0$ in free space are given in the table.
The mass of the
quark $m_u=m_d=5.5$ MeV and $m_s=150$ MeV. The constants are fixed
for Nucleon mass 939 MeV,  $\Theta^+$ mass 1540 MeV and  $\Xi^{--,0}=1862$ MeV. The $\sigma$
and $\omega$ nucleon couplings are fixed from the saturation properties
of nuclear matter i.e. $E_B=15.7$ MeV at $\rho_0=0.15$ fm$^{-3}$}
\begin{ruledtabular}
\begin{tabular}{lcccccc}
$R_N$ (fm) & $g_{\sigma N}^q$ & $g_{\omega N}$ & $B_\Theta^{1/4}$ & $Z_\Theta$& $Z_\Xi$ & \\
\hline
0.6 & 5.957  & 8.981 & 210.854 &6.3765 &  5.4869& all quarks in $s$ state \\
 &  &  & &  7.5359 &  6.6445 & $u$ or $d$-quark in $p$ state \\
 &  &  &  & 7.4991 & 6.6052 & $s$-quark in $p$ state \\
\hline
1.0 & 5.519  & 7.622 & 143.745 & 3.8428 & 2.3919& all quarks in $s$ state \\
 &   &  &  &  4.9969 &  3.5479 & $u$ or $d$-quark in $p$ state \\
 &   &  &  &  4.9298 & 3.4749 &  $s$-quark in  $p$ state \\
\end{tabular}
\end{ruledtabular}
\label{tab1}
\end{table}

\end{document}